\documentclass[12pt]{article}

\usepackage{graphicx}  
\usepackage{dcolumn}
\usepackage{bm}
\usepackage{graphics}
\usepackage{rotating}
\usepackage{authblk}

\newcommand{\p}[1]{\mathrm{#1}}

\begin{document}

\title{Instrumentation for the Energy Frontier}

\author[1]{Ulrich Heintz}
\author[2]{Daniela Bortoletto}
\author[3]{Marcus Hohlmann}
\author[4]{Thomas LeCompte}
\author[5]{Ron Lipton}
\author[1]{Meenakshi Narain}
\author[6]{Andrew White}
\affil[1]{Brown University}
\affil[2]{Purdue University}
\affil[3]{Florida Institute of Technology}
\affil[4]{Argonne National Laboratory}
\affil[5]{Fermilab}
\affil[6]{University of Texas, Austin}

\maketitle

\begin{abstract}
The Instrumentation Frontier was set up as a part of the Snowmass 2013 Community Summer Study to examine the instrumentation R\&D needed to support the particle physics research over the coming decade. This report summarizes the findings of the Energy Frontier subgroup of the Instrumentation Frontier.
\end{abstract}


\section{Introduction}
During the last three decades, we have explored physics at the energy frontier in ever more challenging environments. First at SLC and LEP, then at the Tevatron, and now at the LHC, we have carried out precision measurements of the parameters of the standard model, we have observed the top quark, and most recently, the Higgs boson. These successes have been made possible by advances in technology and instrumentation that allowed us to build ever more complex detectors. In order to continue our success at the energy frontier further advances in technology will be required. To succeed we need to make technical and scientific innovation a priority as a field. 

\section{Physics at the energy frontier}

The standard model provides us with a seemingly complete description of elementary particle physics as we observe it in the laboratory. Yet we know that it is not a complete theory of the physics of our universe. We know that most of the matter in the universe cannot be explained by the standard model and the standard model does not describe gravity. The energy scale of gravity appears to be so fundamentally different from the energy scales at which we have verified the validity of the standard model that it seems unlikely the standard model could be correct over the entire spectrum of energies that determine the fate of our universe. The recent discovery of the Higgs boson has confirmed the last untested prediction of the standard model. Yet the relatively small mass of the Higgs boson is the reason that we believe the standard model cannot be valid to much higher energy scales than we can probe in the laboratory at this time and that there must be new physics. The big question that we want to answer is: what is the nature of this new physics?

Our exploration of particle physics has been characterized by a two pronged approach. On the one hand we have pushed to ever higher energies in order to experimentally probe the structure of the standard model by directly producing particles that cannot be produced at lower energies, such as the top quark and the Higgs boson. On the other hand we have carried out extremely precise measurements of phenomena at lower energies to extract information about the underlying physics. In the cases of both the top quark and the Higgs boson, we had a very good idea at what mass we should expect to find them, based on precision measurements at lower energies. It is the synergy between these two approaches that gives us the confidence that we understand the underlying physics. It is therefore essential that we explore the boundaries of the standard model in both ways. We need to directly look for new massive particles at the energy frontier and dark matter particles at the cosmic frontier as well as carry out ever more precise measurements of lower energy phenomena at the intensity frontier in order to deepen our understanding of the physics that drives our universe. The questions that can be explored at the energy frontier in the coming decades also reflect this two-fold approach. 

One of the most important questions is whether the new 125 GeV boson is the minimal Higgs boson of the standard model or part of an extended Higgs sector. In order to discriminate between these possibilities, we need to measure the properties of the Higgs boson to the best possible precision. This includes measuring the cross sections of its production processes and its branching fractions to determine the complete set of couplings and self-couplings of the Higgs boson. In addition we also want to search for the other particles of a potential extended Higgs sector and we want to study WW scattering at high energies to understand what regularizes its amplitude. 

Furthermore, we want to search for manifestations of the new physics that we believe to exist at the TeV scale. We want to continue our search for superpartners and explore corners of SUSY parameter space that we have not been able to reach. Light stop squarks or compressed spectra could make it very difficult to observe the superpartners that are in our range because their decay products could be relatively soft. Of course we also want to search for new massive particles at as high a mass as we can reach. 

In order to maximize our sensitivity to these we want to push beam energies as high as possible. The study of the Higgs boson and the search for low scale SUSY ensure that, even as the beam energy increases, we are still looking at electroweak scale phenomena involving W and Z bosons and their decay products. The challenge for our detectors will be to maintain acceptance to relatively soft particles, to maintain large angular acceptance to minimize theoretical uncertainties and retain sensitivity to distinguish between different models should we find something new. Superior spatial and time resolution will be essential for pattern recognition in high occupancy environments.

\section{Benchmark processes}

Here we identify specific benchmark processes that define the detector performance requirements.  

Precise measurements of Higgs boson decays define important benchmarks for detector performance. 
At hadron colliders, H$\rightarrow\gamma\gamma$ decays require the ability to trigger on photon pairs with low momentum thresholds. In order to achieve optimal diphoton mass resolution the identification of the hard scattering vertex is important H$\rightarrow$ZZ$\rightarrow4\ell$ and H$\rightarrow$WW$\rightarrow\ell\nu\ell\nu$ decays require the ability to trigger on multiple leptons with low momentum thresholds. Measuring the branching fractions for $\p{H}\to\p{b}\overline{\p{b}}$ and especially $\p{H}\to\p{c}\overline{\p{c}}$ drives the flavor tagging performance of vertex detectors. The latter can probably only be measured at lepton colliders and R\&D is still needed to develop the technology for a vertex detector that achieves the required impact parameter precision of 5~$\mu$m at high $p_T$. The momentum resolution of the tracking system at $\p{e}^+\p{e}^-$ colliders should be $\sigma(1/p)\approx 2.5\times10^{-5}$. This is driven by the Higgs mass measurement using the Z boson recoil.Separating $\p{H}\to\p{WW}$ and $\p{H}\to\p{ZZ}$ decays followed by hadronic decays of the vector bosons drives requirements for the jet energy resolution. At hadron colliders the measurement of Higgs boson production by the vector-boson fusion mechanism requires forward coverage for jets up to $|\eta|\approx4$, including the capability to trigger. Pileup mitigation will be required to be able to veto jets between the two forward/backward jets. The detection of the decay $\p{H}\to\mu\mu$ drives the detector acceptance and efficiency to detect muons.

Searches for physics beyond the standard model also define some benchmarks. The analysis of the LHC data has significantly restricted SUSY models. However there are corners in parameter space for low-mass SUSY that remain. If the mass spectrum of superpartners is compressed their decay products could be rather soft, resulting in relatively low missing transverse momentum and low momentum charged particles. This affects the designs of the trigger and the tracking detector. Triggering on moderate missing $p_T$ requires effective pileup mitigation, possibly using tracking at the first trigger level to compute missing $p_T$. Large acceptance for leptons and triggering on low $p_T$ leptons as a part of more complex trigger conditions will be important. Calorimeter timing could also help mitigate pileup by separating soft particles from the hard scattering vertex and from pileup interactions.

The range of phenomena for more exotic physics models is wide. The capability to flavor tag very high mometum jets (for example from high mass exotic resonances) could be important. For jet $p_T$ values in excess of 500~GeV, the b-decay length becomes comparable to the radius of the inner layer of vertex detectors. Tracks from such secondary vertices will not leave hits in the innermost vertex detector planes and they could be boosted such that they are very close together in space requiring enhanced two-track separation capabilities. Calorimeter mass resolution could be important for separating closely spaced resonances that decay to jets.

One might ask whether there are any advances in technology that could revolutionize the physics capabilities of particle physics detectors. The omnipresent constraint on the technical capability of 
a detector is often cost rather than technology limitations. Therefore develoments of techniques that reduce the cost of the large scale detectors required for high energy experiments by an order of magnitude may be the advances that could have the most profound impact on the high energy physics program in the US. An example is forward tracking for the phase 2 LHC detectors. The technology to build trackers that can measure particles with $|\eta|<4$ exists. But constructing the detectors is costly because of the high density of connections that need to be made to sufficienctly pixelated detectors. A technique that would reduce the cost of such detectors by an order of magnitude would make a big impact. 

A transformational capability for hadron collider detectors would be the ablility to read out the entire detector for every beam crossing and process the data for the first level trigger decision. Much R\&D goes into the direction of making more and better data available for the lowest level trigger, such as the efforts to develop a level 1 track trigger. Crucial elements that support this trend are the development of low-power, high-bandwidth data links and highly interconnected architectures for the trigger processors.

Besides cost one should also keep an open mind for new ideas from other fields. Examples are the use of novel materials such as graphene or metamaterials, advanced photonics techniques, and developments in the telecom industry. 

\section{Challenges for energy frontier machines}
Experiments at the energy frontier are dominated by large accelerators and multipurpose collider detectors. We want these detectors to provide the highest quality information such as: measure position to microns, measure timing to picoseconds, measure energy deposits to eV using highly pixelated trackers and calorimeters and read out all the data for as many events as possible. All this at low cost, using minimal power, and with low mass. Real detector designs will of course be a compromise between the constraints of the accelerator environment, the physics needs, our technical capabilities, and the available funds. 

\subsection{Hadron colliders}

Table \ref{tab:ppcolliders} lists some parameters for current and possible future hadron colliders \cite{ESG}. 

\begin{table}
\centering
\caption{Hadron collider facilities.}
\label{tab:ppcolliders} 
\begin{tabular}{lcccc}
\hline\hline
						& LHC 		& HL-LHC 	& HE-LHC 	& VLHC \\ \hline
time scale 					& 2015-2021 	& 2023-2030 	& $>$2035 	& $>$2035 \\
center of mass energy (TeV) 	& 14			& 14  		& 26-33  		& 100  \\
luminosity (/cm$^2$/s) 		& $10^{34}$ 	& $5\times10^{34}$ & $2\times10^{34}$ &  \\
integrated luminosity (/fb) 		& 300 		& 3000 		& 3000 		& 1000 \\
number interactions/crossing 	& $\approx$50 	& $\approx140$ & $\approx225$ & $\approx40$ \\
bunch spacing (ns) 			& 25  		& 25 		& 50 		& 19 \\
radiation dose (Gy @ R=5 cm) 	& $3\times10^4$  & $5\times10^6$ &  \\
\hline\hline
\end{tabular}
\end{table}

The challenges that experiments at the LHC and other hadron colliders face are high interaction rate, large number of interactions in the same beam crossing, and high radiation doses. The high interaction rate will require an increased rejection power of the trigger system and low power-high bandwidth data links to transport the data out of the detector. Dealing with high pileup requires increased pixelization of the detector and 10-100 ps timing to separate the particles that emanate from different interactions in the same beam crossing. The increased radiation dose requires the development of radiation hard detector technologies and detector operation at increasingly lower temperatures.

\subsection{e$^+$e$^-$ colliders}

Table \ref{tab:eecolliders} lists some parameters for e$^+$e$^-$ colliders. 

\begin{table}[h]
\centering
\caption{e$^+$e$^-$ collider facilities.}
\label{tab:eecolliders} 
\begin{tabular}{lccc}
\hline\hline
 						& ILC 	& CLIC 	& TLEP \\ \hline
timescale					& 2028	& $>$2035 & $>$2035 \\
center of mass energy (GeV) 	& 91-1000	& 350-3000 & 91-350 \\
luminosity (/cm$^2$/s) 		& $5\times10^{34}$ 	& $6\times10^{34}$ & $56\times10^{34}$ \\
integrated luminosity (/fb) 		& 1000	& 3000 	& 2500 \\
bunch spacing (ns) 			& 366 	& 0.5 \\
bunches/train 				& 2625 	& 312 	& 4400 \\
length of bunch train (ms) 		& 1 		& 0.156 	& n/a \\
interval between trains (ms) 	& 199 	& 20 	& n/a \\
collision rate (Hz) 			& 5 		& $<50$ 	& \\
\hline\hline
\end{tabular}
\end{table}

Challenges at e$^+$e$^-$ colliders are different from hadron colliders. Exploiting the physics potential mandates detectors that have higher granularity and lower mass by one to two orders of magnitude than the current LHC detectors. On the other hand requirements imposed by beam structure, data rate, and radiation dose are much more modest than at LHC\cite{ILC}.

An exception to this is the time structure of the beam at CLIC which foresees bunch trains at 20 ms intervals. Each train comprises 312 bunches separated by 0.5 ns. There will be at most one e$^+$e$^-$ interaction per train. However, the high CLIC energies and small intense beams lead to significant beam-induced background. Triggerless readout of zero-suppressed data at the end of the bunch train is planned. Hit time resolution of 1 ns combined with fine detector granularity are essential for efficient suppression of beam-induced background in the data\cite{CLIC}. 

\subsection{$\mu^+\mu^-$ colliders}

The vision for a muon collider includes a number of incremental steps. The first step would be a Higgs factory muon collider running at a center of mass energy equal to the Higgs mass and a luminosity of about $10^{31}-10^{32}$/cm$^2$/s. The beam energy spread at a muon collider is small enough that it can produce Higgs bosons in the s-channel. This could be followed by an energy frontier accelerator with a center of mass energy of 3-6 TeV and a luminosity of  about $10^{34}-10^{35}$/cm$^2$/s. 

At a muon collider the bunch spacing would be relatively large, 500 ns for a Higgs factory and 10 $\mu$s for a high energy collider, leaving lots of time to read out the detector. However, backgrounds are large. For two beams at 0.75 TeV there will be $1.3\times10^{10}$/m/s muon decays. The radiation dose will be $10^3$ -- $10^4$ Gy/year. Thus, detectors must be radiation hard. Precision time resolution will be crucial to separate signals from the background and the primary interaction as most of the background is soft and out of time\cite{MuC,vertexMuC}.

\subsection{Performance goals}

Table~\ref{tab:ppperformance} summarizes the performance figures used for the "snowmass detector" that served as a reference for hadron collider simulations and corresponds to the typical performance of ATLAS and CMS. Table~\ref{tab:eeperformance} summarizes the performance goals for lepton colliders.

\begin{table}[h]
\centering
\caption{Performance highlights of "snowmass" hadron collider detector\cite{BlueTable}.}
\label{tab:ppperformance} 
\begin{tabular}{lc}
\hline\hline
performance figure of merit		& goal \\ \hline
tracking/muon system acceptance 	& $|\eta|<2.5$ \\
calorimeter acceptance	   		& $|\eta|<5$ \\
EM energy resolution (central) $\sigma_E/E$ & $8\%/\sqrt{E}\oplus0.5\%$ \\
jet energy resolution (central) $\sigma_E/E$ & $60\%/\sqrt{E}\oplus3\%$ \\
track momentum resolution & 1.5\% @ 100 GeV \\
\hline\hline
\end{tabular}
\end{table}

\begin{table}[h]
\centering
\caption{Performance goals for e$^+$e$^-$ collider detectors\cite{ILC,CLIC}.}
\label{tab:eeperformance} 
\begin{tabular}{lc}
\hline\hline
performance figure of merit		& goal \\ \hline
track impact parameter resolution $\sigma_b$	& $5\oplus10/p\beta\sin^{3/2}\theta$ \\
track momentum resolution $\sigma_{pT}/p_T^2$ 	& $2\times10^{-5}$/GeV  \\
jet energy resolution $\sigma_E/E$ 		& 3.5\% @ 100 GeV  \\
lepton acceptance 			& $|\theta|>10$~mrad \\
time resolution				& 1-10 ns for CLIC \\
\hline\hline
\end{tabular}
\end{table}

\section{Technology R\&D themes}

Instrumentation R\&D will have to address the challenges posed by the physics goals and the accelerator requirements outlined in the two previous sections. Several themes emerge from these.

\begin {itemize}

\item{Pixelization:} Continued microelectronics feature size reduction, 3D electronics technologies, and new interconnect technologies will allow finer spatial segmentation of detectors. Low power-high bandwidth links need to be developed for transmission of the increased data volume.

\item{Timing:} Faster sensor technologies and low power, fast electronics are needed to enable precise timing measurements for detector signals. Pixelization and 3D technologies will allow the lower capacitance detectors needed and waveform digitization can provide precise timing information.

\item{Resolution:} Increased pixelization of trackers and calorimeters will improve the resolution of tracking and calorimetry. Imaging calorimeters combined with better understanding of hadron showers could enable a transformational improvement in the resolution of jet energy measurements.

\item{Mechanics and power:} New materials and techniques need to be developed to provide mechanical support for thinner, more highly segmented detectors, reduce the power dissipated in increasing numbers of electronics channels, and improve cooling to enable detectors to run at lower temperature. These could include carbon fiber supports, foamed thermally conductive materials, CO$_2$ cooling, and power delivery using DC-DC conversion or serial powering.

\item{Data transmission:} Low power optical interconnects, wireless data transmission, and low power signaling could revolutionize the transmission of data out of the detector.

\item{Cost:} The new technologies must also be more cost effective. Large area arrays utilizing new technologies (as used by LAPPD), wafer scale integration, high yield assembly techniques with active edge sensors could help achieve this goal.

\end {itemize}

We have identified three detector areas for which new technologies would have the biggest impact. These are: trigger and data acquisition, tracking, and hadron calorimetry. We associate the above R\&D themes with these detector areas and summarize the status of R\&D, based on the white papers submitted to the intensity frontier working group. The full list of white papers can be found at 

{\tt https://www-public.slac.stanford.edu/snowmass2013/Index.aspx}.

\subsection{Advanced triggers and data acquisition}
As the instantaneous luminosity at the LHC increases by an order of magnitude, increased pileup leads to non-linear increases in trigger rates. Improved rejection power is needed at each trigger level. Access to the full granularity of the detector at level 1, and a level 1 track trigger provide the data for advanced pattern recognition. Compact high-density optical connectors and low power-high bandwidth optical links\cite{opticalLinks} help transfer this data out of the detector and ATCA and $\mu$TCA crate technology  with high-speed star and mesh backplanes used by the telecommunications industry provide increased interconnectivity between processors\cite{ATCA}. Processing will benefit from 3D technology for associative memory ASICs and exploit state of the art FPGAs and processing units such as GPUs\cite{triggerLHC}.

There are two approaches to designing a L1 track trigger for the LHC experiments. A region-of-interest trigger\cite{ATLAS} would read out only hits near level 1 electron or muon candidates and help in sharpening $p_T$ turn-on curves. This will allow raising the $p_T$ threshold at level 1 and thereby reduce the rate without loss of efficiency for high $p_T$ leptons. The other approach is a self-seeded trigger\cite{L1TT}. This requires on-detector data reduction which can be achieved by using coplanar strip or pixel sensors, closely spaced in radial direction to reject hits from low $p_T$ tracks. A self-seeded track trigger could potentially reconstruct all tracks above a moderate $p_T$ threshold such as 2 GeV and associate them with primary interaction vertices. This would enable track-based isolation using only tracks from the same vertex thereby mitigating the effect of pileup. 

Application specific integrated circuits (ASICs) are fundamental component of instrumentation for all frontiers\cite{ASICs}. They allow high channel density, improve analog performance (e.g. noise, speed), enable data reduction, lower power dissipation, and reduce cabling and mass. R\&D is needed to develop high-speed waveform sampling\cite{waveform}, pico-second timing, low-noise high-dynamic-range amplification and shaping, 
digitization and digital data processing, high-rate data transmission, low temperature operation, and radiation tolerance.

\subsection{Next generation trackers}
Trackers need to provide excellent momentum and impact parameter resolution in high rate environments. This requires improved two track separation. The ability to assign a time stamp to hits will be important to reduce backgrounds especially at CLIC and at muon colliders. Thinner sensors are required for the short charge collection times that enable precise timing measurements. Spatial resolution also is ultimately limited by sensor thickness. Charges spread less in thinner sensors providing better spatial resolution. Thinner sensors also reduce mass, leakage current, and power dissipation. At hadron colliders and muon colliders, radiation hardness is a must for tracker sensors. To mitigate the effects of radiation damage, sensors must be operated at low temperatures. On the other hand, power distribution for the increased channel count and fast data links is an issue. These trackers will require the development of multipurpose structures for mechanical support and more efficient cooling.

Hybrid pixel detectors are the current state-of-the-art for pixel trackers\cite{hybridpixel}. They consist of two separate silicon devices, the sensor and the read out chip, that are bump-bonded together. They can tolerate radiation up to fluences of $5\times10^{15}$ neq/cm$^2$ and operate at rates up to 300 MHz/cm$^2$ using readout chips created with 130nm CMOS technology. For HL-LHC the rate and dose requirements will approximately triple. A readout chip that can handle these requirements would need to use smaller feature sizes (65nm CMOS) and contain some 500 million transistors. The design of such a chip would be a large project that requires international collaboration. The RD53 Collaboration was formed to develop a new pixel chip for ATLAS and CMS. 

In monolithic active pixel sensors (MAPS) sensor and readout circuitry are implanted in the same Si wafer, providing a single chip solution without bump bonding\cite{MAPS}. MAPS have less mass than hybrid pixel detectors, lower capacitance and can be made highly granular and thinned to about 50 $\mu$m thickness. Finer pixelization will improve spatial resolution and two-track separation. At this thickness the sensors no longer dominate the mass in the tracker and cables, support, and cooling services become important. Sensor stitching can further reduce the mass of tracking detectors by routing signals and clocks through metal lines on the chip. 3D integration would allow combination of sensor and readout chip with different feature sizes/technologies\cite{active3D}.

An alternative approach to reducing the effective thickness are 3D pixel sensors\cite{3Dsensors}. Planar sensors collect charge in implants on the sensor surface. 3D sensors collect charge in implant columns in the bulk material. The depletion depth is now equal to the distance between implant columns and can be made much smaller than the sensor thickness. With smaller depletion depth these sensors can collect charge faster, have lower leakage current, lower lower depletion voltage, lower power dissipation, and are more radiation tolerant.

A further step are 4D ultra-fast silicon detectors which combine precise spatial resolution with ps time resolution based on  silicon thinned to $\approx5 \mu$m to reduce charge collection time\cite{4Dsensors}. Sensors this thin do not generate enough charge to guarantee near 100\% hit efficiency. Therefore the charge must be increased using charge multiplication in the bulk of the sensor. This can be achieved by tuning the depth doping profile of the bulk material to create a low resistivity region just below the pixel implant. This creates a high field region in which a moderate charge multiplication gain can be achieved. Readout systems to match sensor rate, segmentation, and time measurement capabilities need to be developed. Significant R\&D is still required in the areas of wafer processing options (n-bulk vs p-bulk, planar vs 3D sensors, epitaxial vs float zone) and depth and lateral doping profile.

Diamond sensors also hold promise for the development of radiation hard detectors. Chemical vapor deposition (CVD) diamond has a band gap of 5.5 eV (silicon: 1.1 eV) and a displacement energy of 42 eV/atom (silicon: 15 eV). Thus, diamond is expected to be intrinsically more resistant to damage by ionizing radiation and does not require extensive cooling. Another attractive features is that diamond is a low Z material. The flip side is that diamond only generates 60\% as many charge carriers as silicon\cite{diamond}.

Some of the above technologies can be combined to develop new techniques to produce large area, low cost pixelated tracking detectors\cite{3Dtechnology}. These utilize wafer-scale 3D electronics and sensor technologies currently being developed in industry. Readout chips and sensors can be connected at the wafer level using through silicon vias and oxide bonding. Active edge technologies reduce dead areas at the sensor edges, allowing the assembly of large detector modules by tiling the bonded chip-sensor assemblies. This can result in fully active sensor/readout chip tiles which can be assembled into large area arrays with good yield and minimal dead area.

Finally, support, cooling, and power are also in need of R\&D. Next generation trackers will have increased power density, high channel count, high speed data links, and be exposed to high radiation doses. They require efficient cooling and low mass support and services. Current R\&D directions are the development of new materials which are low Z, stiff, thermally conductive, and radiation hard. Possible candidates are carbon foams/fibers and ceramics. The trend is towards multifunction structures that combine mechanical support, cooling and power services in one structure\cite{support}. Radiation tolerant DC/DC converters that can operate in magnetic fields are also a current R\&D direction to reduce power dissipation by bringing in power at higher voltage and reducing it on detector to the required chip supply voltages\cite{DCDC}.

\subsection{High resolution hadron calorimeters}
In order to separate hadronic W and Z boson decays a jet energy resolution of  $\sigma \leq 30\%\sqrt{E}$ is required. The resolution of jet energy measurements is limited by fluctuations in the fractions of the total jet energy carried by electrons and photons, neutral hadrons, and charged hadrons. 

One approach to reduce the effect of these fluctuations is to build compensating calorimeters that have the same response for the electromagnetic and hadronic components of the shower. The disadvantage is that the compensation relies on neutrons liberated in the material which is a slow process. Therefore  the signal has to be integrated over a rather long time to achieve the compensating effect which is not compatible with the bunch crossing times at current and future energy frontier facilities. 

Another approach are dual readout calorimeters\cite{DualCal}. These measure both Cerenkov light and scintillation light. The Cerenkov light originates largely from the electromagnetic component of the shower whereas the scintillation light comes from both the hadronic and electromagnetic components. Measuring both components allows the determination of the electromagnetic to hadronic ratio on a jet-by-jet basis. The jet energy can then be corrected depending on this ratio. The theoretical limit for the resolution that can be achieved in this way is $\sigma \leq 15\%\sqrt{E}$. This principle can be implemented with sampling calorimeters, e.g. using Pb/Cu + scintillating fibers, or with homogeneous crystal calorimeters. The latter require a dense and economical material to make a hadron calorimeter feasible\cite{crystalHCAL,ngCrystals}. 

Light can be piped out of the calorimeter to photosensors using optical fibers. A drawback is that these fibers are often radiation soft. The development of radiation hard fibers is therefore necessary. An example for such R\&D is the development of intrinsically radiation hard claddings with low refractive index from nanoporous alumina\cite{radhardFibers}. 

For dual readout efficient UV photodetectors are needed. R\&D is being carried out on Silicon Photomultipliers (SiPM)\cite{SiPM}. These are Geiger-mode avalanche photodiodes. They operate with low power, low voltage, and have low noise. They are compact and have excellent timing resolution. They are insensitive to magnetic fields. As they are made of silicon they are sensitive to radiation and need to be cooled after radiation exposure to keep the leakage current down. GaAs and InGaAs may be alternative, more radiation hard materials to be used as photodetectors. Silicon also has small attenuation length for UV light. Silicon Carbide with a bandgap of 3.2 eV is promising for detecting Cerenkov light.

A different approach is to develop a detector optimized for particle flow to reconstruct individual particles in the shower and apply particle specific corrections. Charged particles can best be measured in the tracker and photons in the electromagnetic calorimeter so that only neutral hadrons have to be measured in the hadron calorimeter. This approach is planned for e$^+$e$^-$ collider detectors. Its efficient implementation requires an imaging calorimeter with high granularity that provides a detailed image of shower\cite{imageCal}.

Micro-pattern gas detectors are a possible technology to achieve a sufficiently finely pixelated calorimeter readout\cite{MPGcal}. These detectors potentially can provide low cost, large area detectors with high granularity that are fast and radiation hard. Examples are plasma panel sensors (PPS) which resemble plasma-TV display panels, modified to detect gas ionization in the individual cells\cite{PPS}, resistive plate chambers (RPC)\cite{RPC}, flat panel microchannels\cite{FPMC}, gas electron multipliers (GEMs), and micromegas.

\subsection{High-rate muon systems}

Muon detection systems typically form the outermost subdetector layer in high-energy physics experiments, often covering areas of thousands of square meters in large-scale experiments. Consequently, economic construction of such subdetector systems is mandatory. At the same time, muon rates and radiation loads that need to be handled by muon systems in hadron collider experiments are increasing, in particular in the forward direction. For example, muon rates around few to tens of kHz/cm$^2$ are expected at the HL-LHC. Micro-Pattern Gas Detectors (MPGDs) such as Gas Electron Multipliers (GEMs) and Micromegas have been proven in small-scale tracking applications to handle rates above MHz/cm$^2$ with spatial resolutions of about 50 microns (for normally incident charged particles) and timing resolutions of a few ns. Recent R\&D efforts by the ATLAS, CMS, and RD51 collaborations have succeeded in scaling up this class of radiation-hard, low-mass detectors to the meter-scale, making MPGDs excellent candidates for robust muon detection at the EF.
 
Development of cost-effective MPGD mass production techniques and establishment of an industry base with MPGD mass production capabilities within the US are of interest to facilitate the application of MPGDs for large muon detector systems at the EF. Additional research on alternative production techniques such as additive manufacturing (“3D printing”) for detection elements and possibly entire detectors, as well as the development of advanced anode patterns to reduce the number of required electronics channels have the potential for reducing the cost of MPGD-based detector systems further \cite{MPGtrack}.


\section{Conclusion}
In order to realize our physics goals, we need to invest in technology R\&D. The challenges at energy frontier facilities will be substantial. Detectors at all facilities will require increased spatial pixelization and time resolution. They will have to deal with increasing data volumes. There are many ideas for instrumentation that can address these challenges. 

We need a stable mechanism that provides funding to develop them into viable technologies. An investment into technology R\&D will carry many dividends in the future. Detector technologies that enable the construction of large-scale detectors at affordable cost will increase the depth and breadth of particle physics research that can be carried out within a certain budget. Support for technology R\&D will enable us to explore particle physics more effectively but it will also increase the benefits of the field of particle physics to society, in the form of  technology spinoffs and the injection of highly trained, technically skilled individuals into the work force.



\end{document}